\documentclass[]{spie}  
\usepackage[]{graphicx}

\title{Critical-angle x-ray transmission grating spectrometer with extended bandpass and resolving power {\Large\textrm{$>$}} 10,000} 

\author{Ralf K. Heilmann,$^1$ Alexander R. Bruccoleri,$^2$ Jeffery Kolodziejczak,$^3$ 
Jessica A. Gaskin,$^3$ Stephen L. O'Dell,$^3$ Ritwik Bhatia,$^4$ and Mark L. Schattenburg,$^1$
\skiplinehalf
$^1$Space Nanotechnology Laboratory, MIT Kavli Institute for Astrophysics and Space Research,\\ 
Massachusetts Institute of Technology, Cambridge, Massachusetts 02139, USA\\
$^2$Izentis LLC, PO Box 397002, Cambridge, MA 02139, USA\\
$^3$Marshall Space Flight Center, Huntsville, AL 35808, USA\\
$^4$Ultratech/Cambridge Nanotech, Waltham, MA 02453, USA
}

\authorinfo{Further author information: Send correspondence to R.K.H. E-mail: ralf@space.mit.edu, URL: http://snl.mit.edu/home/ralf}

\graphicspath{%
    {converted_graphics/}
    {/}
}
  
\begin{document} 
  \maketitle 

\begin{abstract}

A number of high priority subjects in astrophysics can be addressed by a state-of-the-art soft x-ray grating spectrometer, such as the role of Active Galactic Nuclei in galaxy and star formation, characterization of the Warm-Hot Intergalactic Medium and the "missing baryon" problem, characterization of halos around the Milky Way and nearby galaxies, as well as stellar coronae and surrounding winds and disks.  An Explorer-scale, large-area ($>$ 1,000 cm$^2$), high resolving power ($R = \lambda / \Delta \lambda >$ 3,000) soft x-ray grating spectrometer is highly feasible based on Critical-Angle Transmission (CAT) grating technology, even for telescopes with angular resolution of 5-10 arcsec.  Still, significantly higher performance can be provided by a CAT grating spectrometer on an X-ray-Surveyor-type mission.  CAT gratings combine the advantages of blazed reflection gratings (high efficiency, use of higher diffraction orders) with those of conventional transmission gratings (low mass, relaxed alignment tolerances and temperature requirements, transparent at higher energies) with minimal mission resource requirements.  They are high-efficiency blazed transmission gratings that consist of freestanding, ultra-high aspect-ratio grating bars fabricated from silicon-on-insulator (SOI) wafers using advanced anisotropic dry and wet etch techniques.  Blazing is achieved through grazing-incidence reflection off the smooth grating bar sidewalls.  The reflection properties of silicon are well matched to the soft x-ray band, and existing silicon CAT gratings can exceed 30\% absolute diffraction efficiency, with clear paths for further improvement.  Nevertheless, CAT gratings with sidewalls made of higher atomic number elements allow extension of the CAT grating principle to higher energies and larger dispersion angles, thus enabling higher resolving power at shorter wavelengths.  We show x-ray data from CAT gratings coated with a thin layer of platinum using atomic layer deposition, and demonstrate efficient blazing to higher energies and much larger blaze angles than possible with silicon alone.  We also report on measurements of the resolving power of a breadboard CAT grating spectrometer consisting of a Wolter-I slumped-glass focusing mirror pair from Goddard Space Flight Center and CAT gratings, performed at the Marshall Space Flight Center Stray Light Facility.  Measurement of the Al K$_{\alpha}$ doublet in 18$^{\rm th}$ diffraction order shows resolving power $>$ 10,000, based on conservative preliminary analysis.  This demonstrates that currently fabricated CAT gratings are compatible with the most advanced grating spectrometer instrument designs for future soft x-ray spectroscopy missions.

\end{abstract}


\keywords{critical-angle transmission grating, x-ray spectroscopy, blazed transmission grating, soft x-ray, grating spectrometer, high resolving power, atomic layer deposition}

\section{INTRODUCTION}
\label{sect:intro}  

Technology development for critical-angle transmission (CAT) gratings has progressed rapidly and now stands at Technology Readiness Level (TRL) 4.\cite{PCOS}  X-ray grating spectrometer (XGS) instruments based on CAT gratings\cite{SPIE10} will provide large effective area and resolving power $R = \lambda / \Delta \lambda >$ 10,000.  CAT gratings already provide high enough diffraction efficiency to meet the requirements of soft x-ray spectroscopy Explorer-size missions such as Arcus,\cite{Arcus} and further gains in efficiency are possible with straightforward fabrication development.  In addition, $R >$ 10,000 has now been demonstrated experimentally with CAT gratings at 1.49 keV.  CAT gratings are prime candidates for insertion in proposed and currently studied missions, such as Arcus or X-ray Surveyor.\cite{surveyor2}

The soft x-ray band contains the characteristic lines of highly ionized carbon, nitrogen, oxygen, neon and iron that are central to studies of the Warm Hot Intergalactic Medium, the search for the missing baryons, the study of the outflows of supermassive black holes and the properties of galaxy halos.  Soft x-ray plasma diagnostics also provide crucial information about the evolution of large-scale structure and cosmic feedback. In addition, soft x-ray spectroscopy of individual stars can reveal the effects of rotation, magnetic fields and stellar winds in stellar coronae.\cite{AEGIS}
Several of these high-priority science questions were identified in the ``New Worlds New Horizons" (NWNH) Decadal Survey,\cite{NWNH} as well as in the 2013 ``Enduring Quests, Daring Visions" visionary Astrophysics Roadmap.\cite{roadmap} 

The HETGS\cite{crc} on Chandra and the RGS\cite{XMM} on XMM-Newton are x-ray grating spectrometer (XGS) instruments that opened the door to soft x-ray spectroscopy with spectral resolving powers much greater than CCD energy resolution.  However, they are based on technology from a generation ago and lack in effective area and resolving power to address the above science questions.  For future XGS instruments high collecting area ($> 1000$ cm$^2$) and high resolving power ($R = \lambda / \Delta \lambda > 3000$) need to go hand in hand - one is of limited use without the other.  An XGS based on CAT gratings\cite{SPIE10} can provide $R >$ 10,000 for even better resolution of kinematically broadened lines, which has now been shown through ray-tracing of Arcus and X-ray Surveyor-type mission concepts,\cite{moritz} and through experimental demonstrations described below.

In preparation for the 2020 Decadal Survey, which will be tasked to prioritize large missions to follow JWST and WFIRST, NASA plans to provide information on several large mission concepts.  One of these mission concepts is the so-called X-ray Surveyor from the visionary roadmap.\cite{roadmap}  An informal X-ray Surveyor mission concept team recently has put together a baseline mission concept assuming a telescope with sub-arcsecond angular resolution and square-meter collecting area.  A microcalorimeter ($\Delta E < 5$ eV) and a wide field-of-view, small pixel imager are envisioned at the focus, and a soft x-ray grating spectrometer with a retractable grating array can provide $R$ on the order of 5,000 simultaneously with an effective area of $\sim 4,000$ cm$^2$, using a single readout array in the focal plane.\cite{surveyor2}  Further subaperturing will boost $R$ well above 10,000.\cite{moritz}  Due to the high transparency of transmission gratings at higher energies a CATXGS will work in perfect synergy with a microcalorimeter at the focus, allowing for simultaneous high spectral resolution observations over the 0.2 - 10 keV range with maximum collecting area.

X-ray transmission gratings offer low mass and relaxed, easily achievable alignment requirements\cite{moritz} compared to reflection gratings.  These properties alone minimize resource requirements due to lower mass support structures and increased tolerance for temperature variations.  Reduced time and effort for alignment in combination with the 15-30 times smaller number of gratings required to cover a given aperture reduces cost. In addition, CAT gratings provide high diffraction efficiencies and can be blazed in higher orders, comparable to the most advanced reflection grating designs.

We briefly describe the CAT grating principle below, followed by an update on fabrication process development.  We then present new results on metal-coated CAT gratings and preliminary results of resolving power measurements.

\section{CAT GRATING PRINCIPLE}

   \begin{figure}
   \begin{center}
   \begin{tabular}{c}

   \includegraphics[height=6cm]{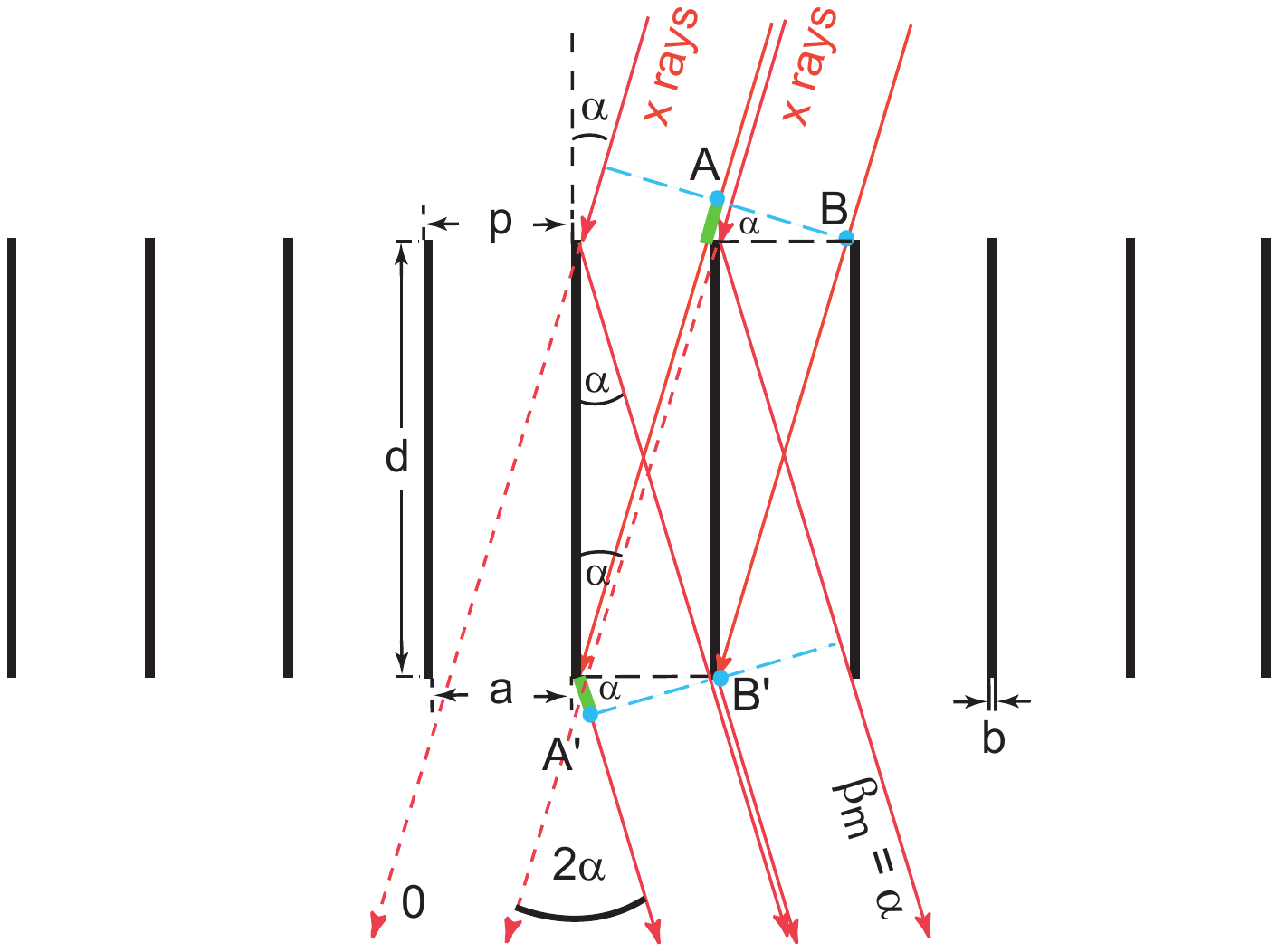}
   \end{tabular}
   \end{center}
   \caption[holder] 
   { \label{fig:CAT} 
     Schematic cross section through a CAT grating of period $p$.  The $m^{\rm th}$ diffraction order occurs at an angle $\beta_m$ where the path length difference between AA' and BB' is $m\lambda$.  Shown is the case where $\beta_m$ coincides with the direction of specular reflection from the grating bar side walls ($\beta_m = \alpha$), i.e., blazing in the $m^{\rm th}$order.}
   \end{figure} 

Critical-angle transmission (CAT) gratings are freestanding transmission gratings with ultra-high aspect-ratio grating bars.  They can be described as blazed transmission gratings and combine the advantages of past-generation transmission and blazed reflection gratings.\cite{CATRFI,OE,SPIE08,AO}
The basic structure of a CAT grating is shown in Fig.~\ref{fig:CAT} in cross section.  X rays are incident onto the nm-smooth side walls of thin, ultra-high aspect-ratio grating bars at an angle $\alpha$ below the critical angle for total external reflection, $\theta_c$ (e.g., $\theta_c = 1.7^{\circ}$ for 1 keV photons reflecting off a silicon surface).  For optimum efficiency the grating depth $d$ should be close to $a/\tan \alpha $ ($a$ being the spacing between two adjacent grating bars), the grating bar thickness $b$ should be as small as possible, and the gratings should be freestanding.  As with any transmission grating, diffraction orders appear at angles given by the grating equation

\begin{equation}
{m \lambda \over p} = \sin \alpha - \sin \beta_m ,
\label{ge}
\end{equation}

\noindent
with $m = 0, \pm 1, \pm 2,...$, $\lambda$ the x-ray wavelength, and $p$ the grating period.  Diffraction orders within a certain angular range around the direction of specular reflection off the sidewalls have enhanced efficiency (``blazing"), as long as $\alpha < \theta_c(\lambda)$.

\section{CAT GRATING FABRICATION}

   \begin{figure}
   \begin{center}
   \begin{tabular}{c}
   \includegraphics[height=7cm]{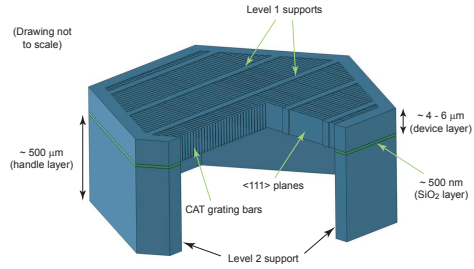}
   \end{tabular}
   \end{center}
   \caption[holder] 
   { \label{fig:unit} 
     Schematic of a grating membrane ``unit cell" (not to scale), formed by a single L2 support mesh hexagon.  The L2 mesh is etched out of the SOI handle layer (back side).  The device layer contains the fine-period CAT grating bars and in the perpendicular direction the coarse, low duty cycle integrated L1 support mesh.  Device and handle layers are separated by the thin buried silicon oxide layer that serves as an etch stop for both front and back side etches.}
   \end{figure} 

We fabricate CAT gratings from $<$110$>$ silicon-on-insulator (SOI) wafers.  The grating bar mask is patterned using interference lithography, with the grating bars aligned parallel to the vertical silicon \{111\} planes.  The CAT gratings and an integrated cross-support mesh (Level 1 or L1 supports) are etched out of the device layer or front side ($\sim 4$-6 micron deep) using deep reactive-ion etching (DRIE).  The device layer thickness determines the depth of the grating.  We use the much thicker SOI handle layer (back side) to separately etch out a high-throughput hexagonal Level 2 (L2) mesh that gives the thin device layer the necessary mechanical strength for a large-area membrane.  Fig.~\ref{fig:unit} gives a schematic representation of the structural hierarchy.

The etched front side is protected with photoresist during the back side DRIE.  After the backside etch the front side is wet-etched in KOH solution, polishing the rough grating bar sidewalls using the \{111\} planes as quasi-etch stops.  In the end a vapor HF etch removes the buried oxide (BOX) in the areas between the L2 supports, creating freestanding gratings.

In the last two years we have made numerous important improvements to our fabrication process.  The KOH step etches the tops of the bars faster than the bottoms.  Through fine-tuning of the DRIE we can now produce bars that are slightly thinner towards the bottom of the device layer, resulting in straight grating bars after the KOH etch. We also introduced improved photoresist protection and BOX removal steps, as well as numerous smaller process improvements.  More detailed descriptions can be found elsewhere.\cite{ahn,ahn2,pran,alex,SPIEalex,alex2,SPIE15,EIPBN2016}

\section{X-RAY MEASUREMENTS}

We performed all of our x-ray diffraction efficiency measurements at beam line 6.3.2 of the Advanced Light Source (ALS) synchrotron at Lawrence Berkeley National Laboratory (LBNL).  A monochromatic beam is incident on the sample at the center of a goniometer.  A slit-covered photodiode mounted to an arm of the goniometer measures the transmitted intensity as a function of diffraction angle relative to the straight-through ($0^{\rm th}$ order) beam.  The gratings are oriented such that the dispersion axis is perpendicular to the horizontal plane of the synchrotron (i.e., the grating bars run horizontally).  We find the grating membrane surface normal through rotation of the grating around the horizontal axis in the grating membrane until positive and negative diffraction peaks of the same order have the same intensity.  (Alternatively, we leave the detector on the $0^{\rm th}$ order, rotate the grating back and forth, and find the angle of symmetry of the measured curve.)  The grating is then rotated around the same axis by some desired angle $\alpha$, resulting in blazing of the diffracted orders that are near the angle $2 \alpha$ from the $0^{\rm th}$ transmitted order.  Orders within a ``blaze envelope" (centered near $2\alpha$) have enhanced diffraction intensity, where the width of this envelope is proportional to $\lambda /a$.

\begin{figure}
   \begin{center}
   \begin{tabular}{c}
   \includegraphics[height=6cm]{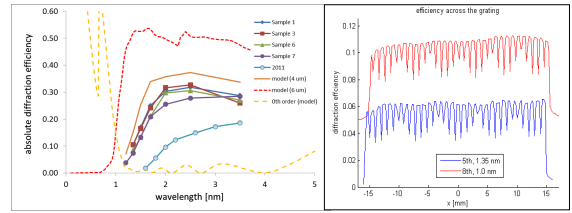}
   \end{tabular}
   \end{center}
   \caption[holder] 
   { \label{fig:effic2} 
     Left: Sum of absolute diffraction efficiencies from orders under the blaze envelope.  ``2011" refers to the the best sample produced via wet etch alone with $\alpha \sim 2.6$ deg.\cite{AO} 
Sample 1-Sample 7 refers to recent samples with $\alpha = 1.91$ deg.  ``Model (4 um)" refers to the theoretical efficiency for the same orders for a perfect CAT grating with similar geometrical parameters as samples S1-S7, also for $\alpha = 1.91$ deg.  ``Model (6 um)" is for a perfect grating with 6 $\mu$m depth and 18\% blockage from L1 supports for $\alpha = 1.5$ deg.
Right: Diffraction efficiency in a given blazed order as a function of location on the grating.  The red curve is shifted for clarity.  Efficiency is fairly constant across the whole sample.  Dips are due to blockage by L2 supports.}
   \end{figure} 

Fig.~\ref{fig:effic2} shows the measured sum of efficiencies of the orders under the blaze envelope for four CAT gratings from last year, and one older sample.  Measurements and theoretical models already include losses due to absorption by L1 supports.  Also shown are the potential gain from going to deeper gratings and the expected efficiency in $0^{\rm th}$ order, available to an imaging instrument at the telescope focus.  The red curve assumes 18\% blockage from L1 supports (a value for structures that we have routinely fabricated).  Our goal in the future is to reduce this number to no more than 10\%, which would move the red curve into the 55\% efficiency range.  

Our recent samples are very homogeneous across the whole grating area in terms of their diffraction efficiency.  This point is demonstrated on the right side of Fig.~\ref{fig:effic2}.  It shows measurements of efficiency with the detector centered on a blazed order, and the sample being scanned through the beam along its surface.

\section{METAL-COATED, EXTENDED BANDPASS CAT GRATINGS}

\begin{figure}[tbp] 
  \centering
  \includegraphics[bb=0 0 657 224,width=5.86in,height=2in,keepaspectratio]{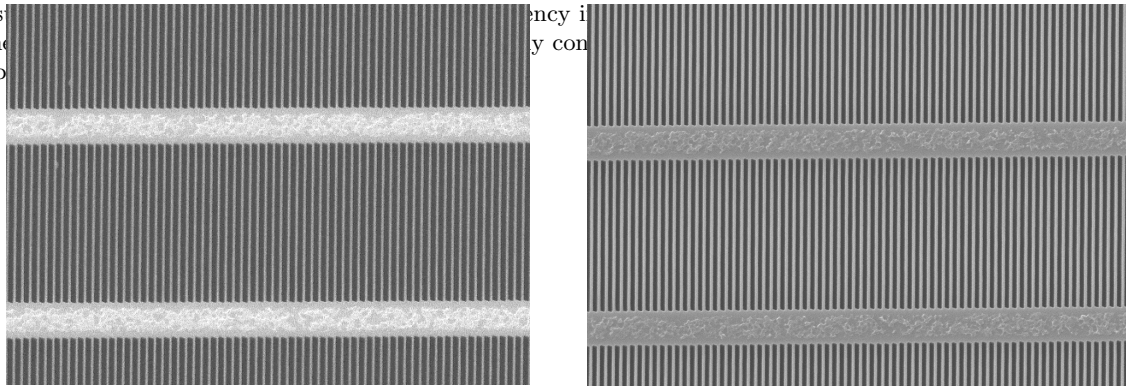}
  \caption{Top down scanning electron micrographs of a freestanding CAT grating, showing the narrow 200 nm-period grating bars and the 5 micron-period L1 cross support mesh.  {\it Left:} Bare silicon CAT grating.  {\it Right:} The same grating after atomic layer deposition of nominally 2 nm  of Al$_2$O$_3$ and 7 nm of Pt.  The bars are coated conformally, the gaps between them are clear, and the increase in bar thickness is hardly discernible on this scale.}
  \label{fig:X3prepostALD}

\end{figure}

\begin{figure}[tbp] 
  \centering
  
  
\includegraphics[bb=0 0 982 386,width=5.67in,height=2.23in,keepaspectratio]{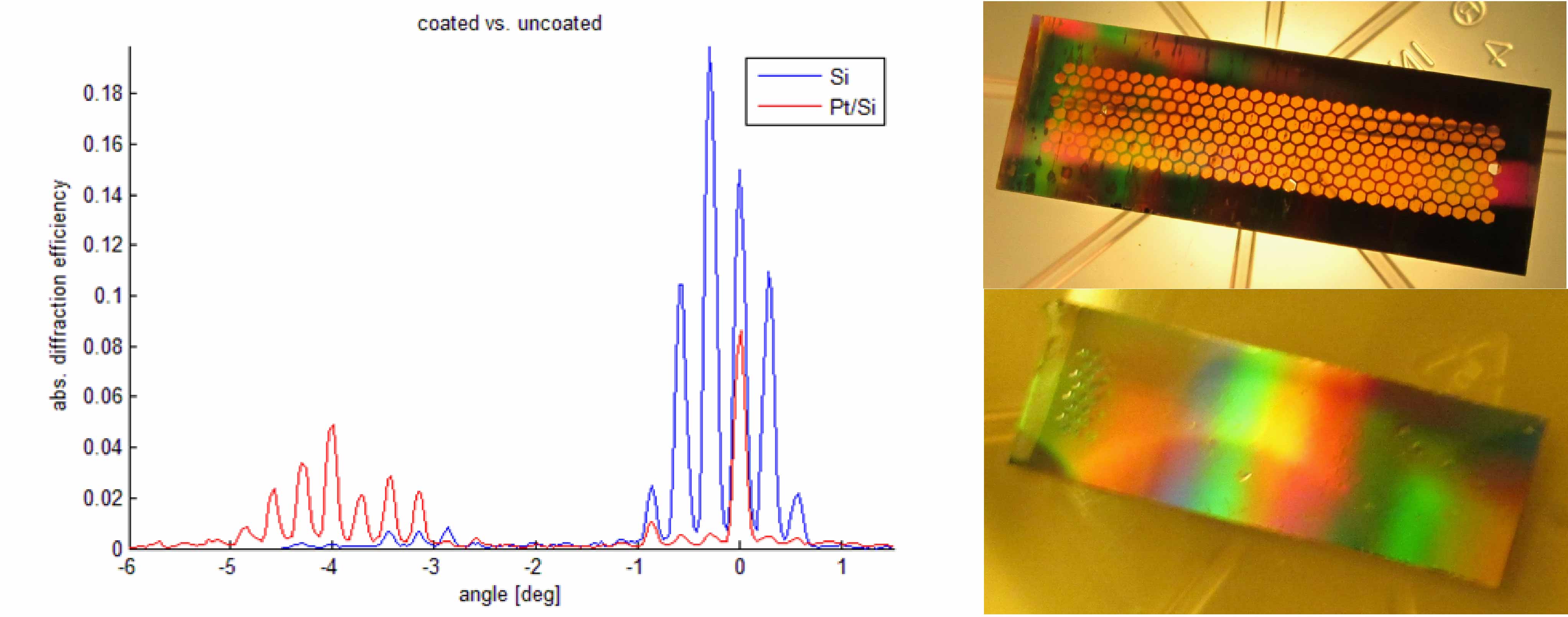}
  
\caption{{\it Left:} Measured diffraction efficiency for a silicon (blue) and a Pt ALD-coated (red) CAT grating for 1.0 nm wavelength x rays, incident at 2 degrees from normal.  This angle is greater than the critical angle for Si at this wavelength, and the grating acts as a phase-shifting transmission grating, diffracting primarily into 0$^{\rm th}$ and surrounding small order numbers.  The incidence angle is smaller than the critical angle for Pt, however, and pronounced blazing takes place around orders near $2\times 2$ deg or 14$^{\rm th}$ order.  {\it Right:} Photograph of a bare Si CAT grating (top) and a Pt ALD-coated CAT grating (bottom).  Visible light diffraction is due to the L1 mesh.  The hexagons cover $\sim 32 \times 9$ mm$^2$ in area.}
  
\label{fig:Pt-Si}

\end{figure}

CAT gratings rely on efficient grazing-incidence reflection of soft x rays off the grating bar sidewalls at
 angles below the critical angle for total external reflection.  Silicon is well matched to the soft x-ray band.  For example, $\theta_c = 1.7^{\circ}$ for 1 keV photons.  However, there are situations where it might be advantageous to be able to efficiently blaze shorter wavelength x rays, or to blaze to larger angles than the critical angle for silicon.  It is well known from x-ray mirrors that coating a material with a small electron density with a very thin film of high electron density material can change the surface reflectivity significantly.  The same is true for the grating bar sidewalls of CAT gratings.  The challenge is how to uniformly coat those sidewalls inside the narrow, ultra-high aspect-ratio gaps between bars.

Atomic layer deposition (ALD) is a proven technique for layer-by-layer growth of metals and dielectrics, even in more extreme geometries than those presented by CAT gratings.  We coated silicon CAT gratings with nominally 2 nm of aluminum oxide as a buffer layer, followed by 5-7 nm of platinum (see Fig.~\ref{fig:X3prepostALD}).  Both steps were done using ALD at $300^{\circ}$C.  The CAT gratings survived ALD at this temperature without a problem - another testament to their robustness.  Potential stress in the ALD layers is not expected to be an issue, since ALD typically leads to conformal coatings on all exposed and chemically identical surfaces.  Any potential thin film stresses are expected to balance each other out since all surface are coated.

We tested several Pt-coated CAT gratings at the ALS.  As seen in Fig.~\ref{fig:Pt-Si}, the metal coating achieved what theory predicts: an increase in the critical angle, and the ability to blaze shorter wavelength x rays and/or to blaze towards larger angles/higher diffraction orders.

A potential drawback of ALD coating is that the space between grating bars gets narrowed.  This could be counteracted through chemical thinning of the silicon bars before ALD.

\section{MEASUREMENT OF RESOLVING POWER}

Measuring high resolving power much greater than 1,000 in a soft XGS is made difficult by several factors.  (1) Polychromatic x-ray optics generally rely on grazing incidence reflection, which leads to long focal lengths and long vacuum beamlines.  (2) It is difficult to diffract soft x rays efficiently by large angles.  (3) Apart from soft x-ray synchrotron beamlines with very sophisticated optics designed for highest spectral purity, there are no natural sources with isolated lines with $\lambda /\Delta \lambda >> 1,000$.  One of the best facilities available for these types of measurements is the Stray Light Facility at the NASA Marshall Space Flight Center.  It features an electron bombardment source on one end of a 92 m beam line, and a large 12 m long chamber on the other end with mounting stages for focusing optics and gratings, and a CCD x-ray detector with a large vertical and horizontal range and focus translation.  We used an Al target at the source that generates characteristic Al K$_{\alpha 1,2}$ lines at 0.83395 and 0.83418 nm wavelengths or 1.4867 and 1.4863 keV in energy.\cite{Schwepp}  The natural widths of these lines are given as 0.241 pm (0.43 eV) in the literature,\cite{OlKr} i.e. the lines are separated roughly by their widths, resulting in significant overlap between them.

\begin{figure}
   \begin{center}
   \begin{tabular}{c}
   \includegraphics[height=10cm]{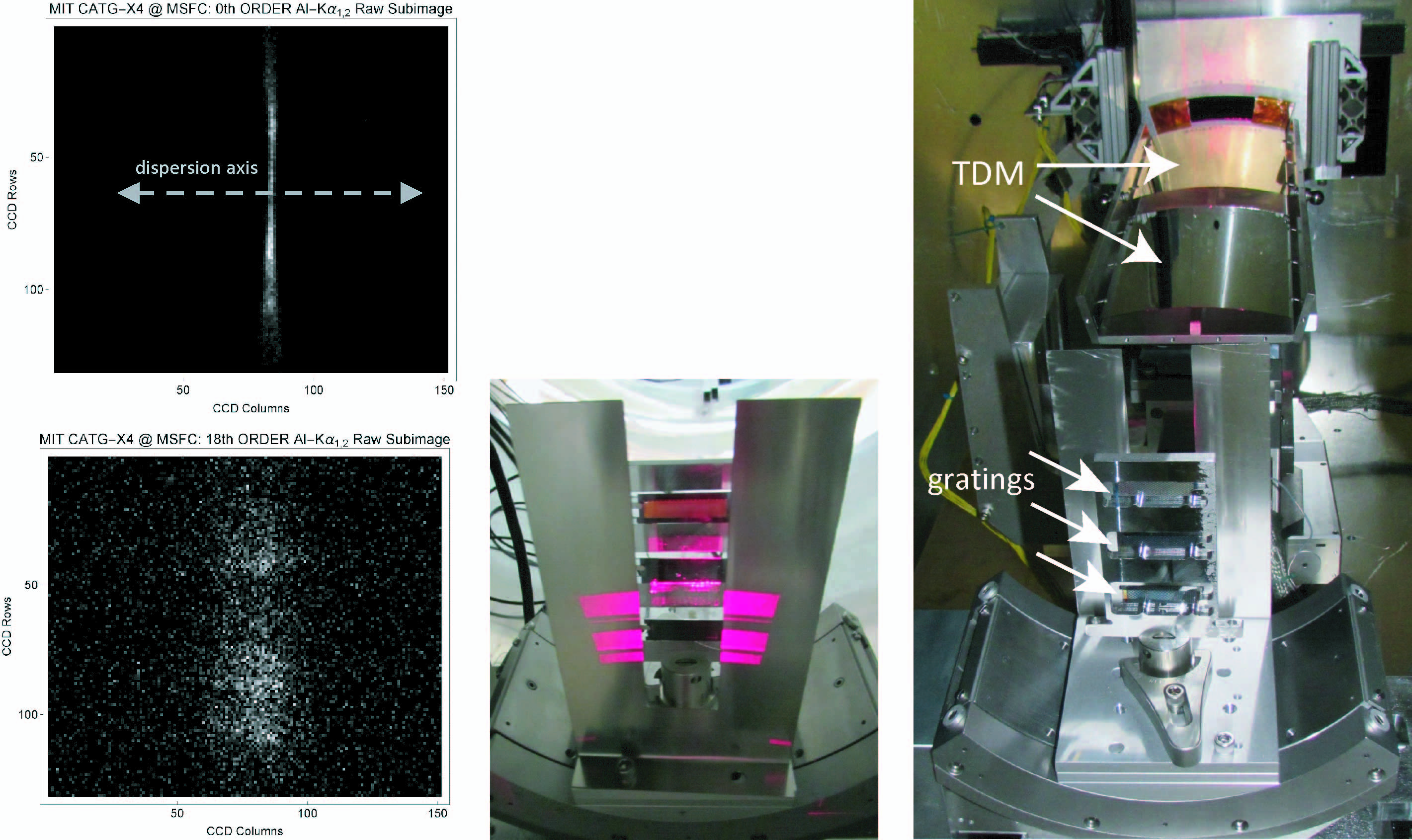}
   \end{tabular}
   \end{center}
   \caption[holder] 
   { \label{fig:MSFC} 
     {\it Left, top:} CCD image of the 0$^{\rm th}$ order beam at focus, transmitted through a CAT grating.  The PSF is the convolution of the mirror PSF (highly anisotropic due to the sub-aperturing effect) and the slit-limited source size.  The gratings are oriented such that the dispersion axis is parallel to the narrow direction of the PSF, minimizing the width of the 1D LSF, which is the projection of the PSF onto the dispersion axis.  The LSF has a FWHM of $\sim 1.2$ arcsec.  {\it Left, bottom:} CCD image of the 18$^{\rm th}$ diffraction order for Al K$_{\alpha}$ radiation.  The doublet structure is already visible by eye.  {\it Center:} Downstream view of the grating mask and the plate holding three gratings during optical alignment.  {\it Right:} Upstream view of the grating plate, grating mask, and hyperbola and parabola segments of the mirror TDM inside the MSFC Stray Light Facility.}
   \end{figure} 

As a focusing optic we used a slumped-glass mirror Technology Development Module (TDM) on loan from W. Zhang's x-ray optics group at Goddard Space Flight Center (GSFC).  It features a parabola-hyperbola grazing-incidence mirror pair in Wolter I geometry with a radius of $\sim$ 245 mm and 8.4 m focal length.\cite{TDM}  The TDM is a sub-apertured optic and provides a highly anisotropic PSF (wider in the plane of reflection, narrow in the perpendicular direction).  A blocking plate with an aperture just upstream of the TDM prevents x rays from bypassing the TDM.  The TDM was placed above its optical axis, and so the plane of reflection was oriented vertically (see Fig.~\ref{fig:MSFC}).

\begin{figure}[tbp] 
  \centering
  
  
\includegraphics[bb=0 0 1200 217,width=6.5in,keepaspectratio]{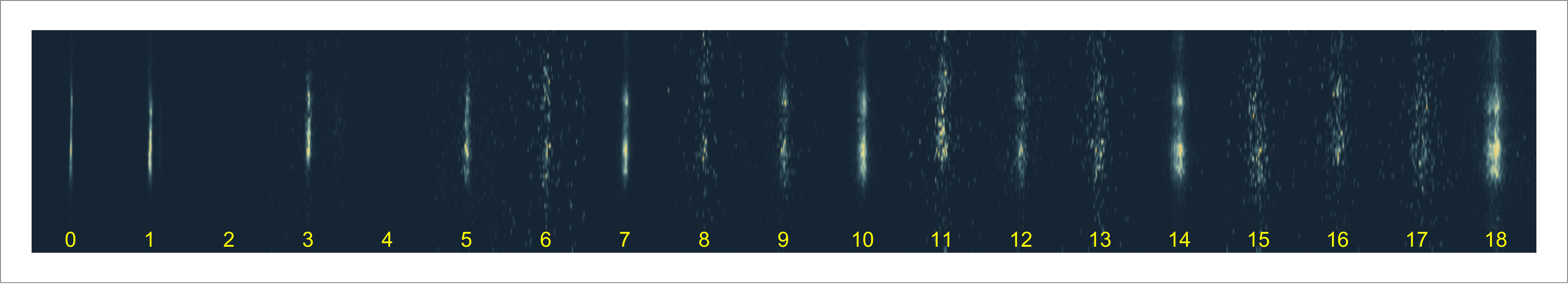}
  \caption{CCD images of orders 0 through 18.  For each order the grating was rotated by half the angle of the diffraction order from normal incidence for optimized blazing.  Integration times varied.  (Orders two and four were not measured.)}
  \label{fig:orders}

\end{figure}

The size of the source can be reduced by a choice of slits (150 or 100 micron wide) near the source, with the slit aligned along the TDM in-plane reflection direction.  This setup produced a focus about 9.25 m from the TDM mirror node, with a 1D line-spread function (LSF) projected onto the out-of-plane axis close to 1 arcsec (see Fig.~\ref{fig:MSFC}).  Roughly 50 cm downstream of the mirror node we placed three 32 mm-wide CAT gratings mounted above each other to a plate.  The plate was mounted to a vertical linear stage on top of a rotation stage (blaze rotation).  A grating mask placed between the TDM and the gratings limited the grating illumination to 30 mm in the horizontal direction, corresponding to a $\sim$ 4 deg azimuthal sub-aperture angle.  Only one grating is illuminated by x rays at a time.  The illumination is in the shape of a wide shallow arc with $\sim$ 1 mm extent in the vertical direction.

We determined the normal-incidence direction for the gratings by eye using an upstream alignment laser.  In order to align the grating dispersion direction to the narrow axis of the TDM PSF we rolled the grating around its normal manually, using visible light diffraction from the L1 support mesh as a guide to the eye.  These two simple steps completed the alignment process.  Transmission gratings can be placed at any reasonable distance from focus; therefore no alignment along the optical axis was necessary.

The straight-through beam is a convolution between the slit-limited source size and the PSF of the TDM, sub-apertured by the grating mask (see Fig.~\ref{fig:MSFC} top left).
To measure the diffracted Al K$_{\alpha}$ spectrum in a given order we rotated the grating by half the diffraction angle of that order for maximum efficiency from blazing.  The measured spectrum is a convolution of the $0^{\rm th}$ order beam, the actual spectrum, and any additional broadening terms, such as period variations in the gratings or aberrations inherent in the optical design.  The diffracted spectrum gets broader in linear fashion with increasing diffraction order, while the direct beam contribution to the measured spectrum remains practically constant.  Therefore we obtain increasingly better resolution with increasing diffraction orders.  We followed this process to $18^{\rm th}$ order, which corresponds to an angle of $\sim$ 4.3 deg.~or 657 mm from $0^{\rm th}$ order (see Fig.~\ref{fig:orders}).  At this angle a LSF of 1 arcsec allows a theoretical maximum resolving power up to $4.3^{\circ}/1$" = 15,480.  The $19^{\rm th}$ diffraction order was near the end of the detector travel and suffered from x-ray leakage around the blocking plate.  (For comparison, the dispersion of the high-energy transmission gratings on Chandra's HETGS is equal to the first-order peak in Fig.~\ref{fig:orders}.)

We performed preliminary data analysis the following way: after background subtraction x-ray counts are projected onto the dispersion axis to produce a 1-D spectrum.  We then try to fit the measured spectrum to a model consisting of two Lorentzians with their natural widths and separation taken from the literature, convolved with the measured $0^{\rm th}$ order LSF.  The result of the first convolution is then convolved with a Gaussian attributed to any additional broadening terms.

\begin{figure}[tbp] 
  \centering
  
\includegraphics[bb=0 0 594 292,width=6.11in,height=3in,keepaspectratio]{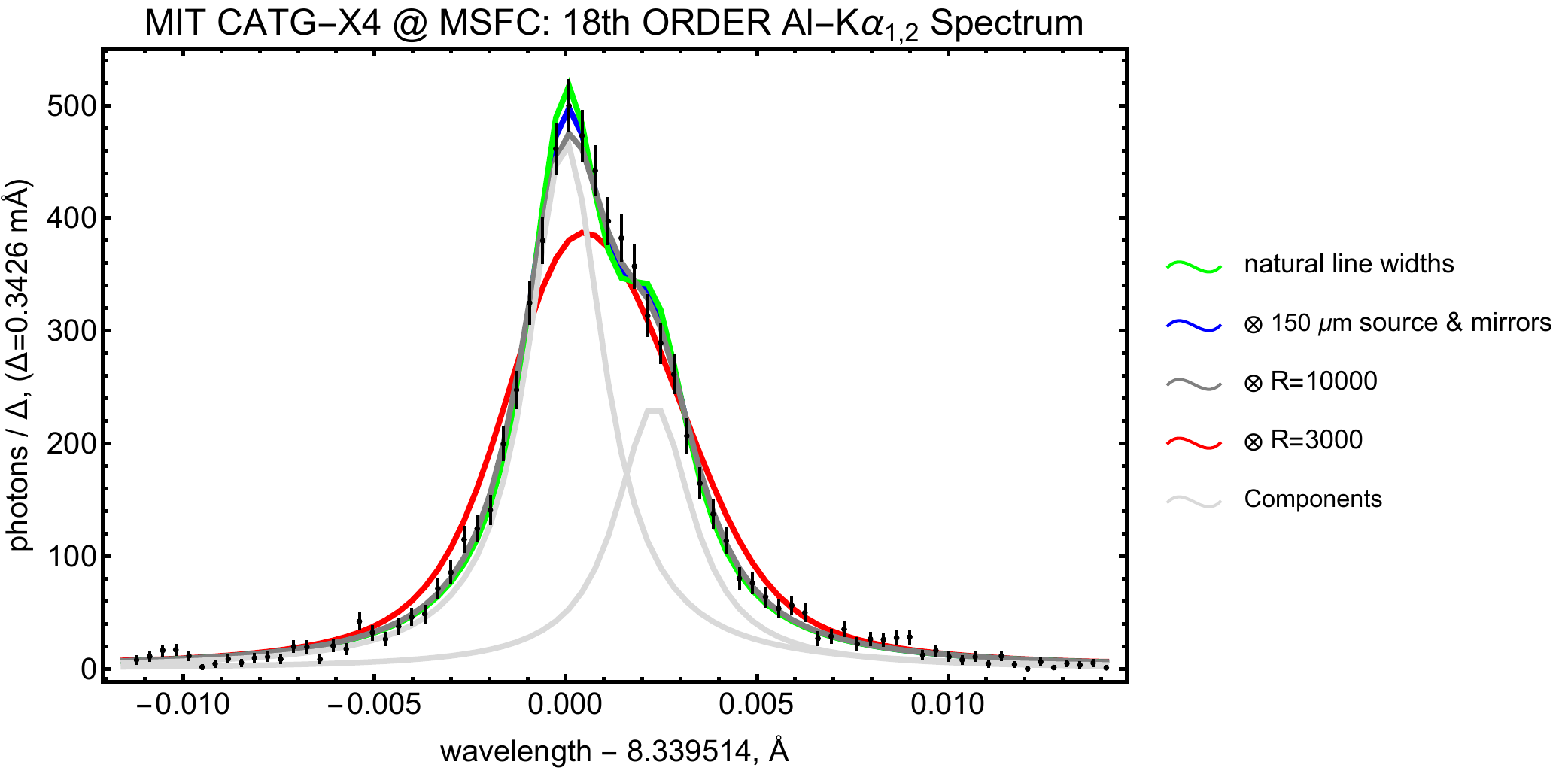}
  \caption{Measured Al K$\alpha_{1,2}$ spectrum in 18$^{\rm th}$ order with 0.15 mm slit (projection onto the dispersion axis) after 4 hour integration.  Vertical lines are the error bars around the number of photons in a given bin.  Light grey curves show the two K$_{\alpha}$ components with their natural Lorentzian widths; the green curve is their sum.  Blue is the convolution of the green curve with the measured source/mirror 1D LSF.  Dark gray is the blue curve broadened due to a Gaussian distribution of grating periods of width $\Delta p/p$ that would limit resolving power to $R = 10,000$.  It falls short at the top, indicating $R >$ 10,000.  The red curve assumes a Gaussian with $R = 3,000$, clearly underestimating $R$.}
  
\label{fig:X418th}

\end{figure}

Fig.~\ref{fig:X418th} compares the measured spectrum with two models that assume that any additional broadening is due to period variations in the gratings, limiting $R$ to 3,000 (red) and 10,000 (dark grey).  The convolution of the presumed actual spectrum with just the measured direct beam already provides a very good fit to the data.  $R = 3,000$ does not fit the data well, but even $R = 10,000$ underestimates the peak in the measured spectrum.  This result shows that this experimental CATXGS setup has a resolving power $>$ 10,000, close to the theoretical maximum for this configuration, and that existing CAT gratings can support spectrometer designs with $R >$ 10,000.

\section{VIBRATION TESTING}

We have performed preliminary vibration tests on two $\sim 32 \times 10$ mm$^2$ CAT gratings glued to aluminum frames, using a shake table and General Environmental Verification Specification (GEVS) parameters.  We performed inspection by eye, optical microscope, and electron microscope before and after the tests and found no structural damage.  While these tests were not comprehensive and did not include before-and-after x-ray tests, the results are positive nevertheless.  We do not expect CAT gratings of our current design to be susceptible to damage from launch vibrations due to their exceedingly high resonance frequencies.  More comprehensive testing will be performed in the future.

\section{DISCUSSION}

CAT grating technology has made major advances in the last two years.  We understand the fabrication process well and are able to control many parameters with good precision.  Working with slightly thicker grating bars ($\sim 60$-70 nm) provides good strength for many processing steps and leads to good yields.  We are routinely producing high-quality CAT gratings $\sim 250$ mm$^2$ in size with good homogeneity, and we do not see any obstacles to significant size increases.

CATXGS effective area is determined by the effective area of the feeding mirror array, the area of the CAT gratings, grating efficiency, and losses due to blockage from support structures.  Grating efficiency can be improved by increasing the grating depth.  We have etched $\sim 6$ micron-deep gratings using KOH or DRIE in the past.  With the valuable experience gained from 4 micron-deep gratings we are confident that we can achieve high-quality 6 micron-deep CAT gratings with our current combined DRIE/KOH process, potentially boosting diffraction efficiency above 50\%.  Further increases may be possible by thinning the grating bars post-fabrication, using repeated cycles of thermal oxidation and vapor HF oxide removal.  We believe that the widths of current L1 and L2 supports can be reduced significantly, leading to less blockage and increased effective area.

ALD coating of silicon CAT gratings opens up a whole new design space for a CATXGS and other applications.  For example, a CATXGS could be envisioned that contains both bare silicon and Pt coated gratings, with effective area as a function of energy fine-tuned for specific science cases.

Our experimental demonstration of $R >$ 10,000 confirms what back-of-the envelope estimates and ray-trace modeling have long predicted: high-quality alignment-insensitive transmission gratings can easily support XGS designs with $R >$ 10,000.

\subsection{CAT Gratings and Arcus}

Arcus is a high-resolution soft x-ray spectroscopy mission to be proposed to the NASA Explorer program.  Its mission performance goals are $>$ 500 cm$^2$ effective area and spectral resolving power up to $R = 3,000$.  Its baseline design uses silicon pore optic (SPO) modules from ESA's Athena technology development efforts as focusing mirrors,\cite{SPO} with a total mirror effective area of about 3,000 cm$^2$, split into four quasi-rectangular petals that span an azimuthal range of about 25 (outer edge) to 65 (inner edge) degrees.  It is conceivable that we can fabricate CAT gratings that cover a complete SPO module each, which would require 152 lightweight CAT gratings with relaxed alignment and flatness tolerances, compared to roughly 2,000 off-plane reflection gratings\cite{randy} that need to meet demanding flatness specifications and more challenging alignment requirements.  We estimate a mass savings on the order of 50 kg or more compared to reflection gratings, or about 5\% of the total spacecraft mass.  CAT gratings also demand less (practically no) power for temperature control.\cite{SPIE10}  CAT grating facets could follow the outline of each SPO module, which would eliminate additional blockage (apart from L1 and L2 supports) from support structures and geometrical mismatch between flat reflection grating plates and curved SPO reflectors.  If we imagine placing today's CAT gratings (32\% efficiency including L1 supports, 19\% blockage from L2 supports) into an Arcus-like mission we would obtain $>$ 750 cm$^2$ peak effective area.  Even after detector losses this number already meets Arcus requirements comfortably.  Improvements in efficiency and throughput outlined above could realistically lead to an effective area on the order of 1,000 cm$^2$ in the very near future.

We have begun to study resolving power for a CATXGS on an Arcus-like mission.  Using SPO-like performance with modules that populate a 30 degree azimuthal wedge we find efficiency-weighted resolving power above 10,000, with relaxed alignment tolerances.\cite{moritz}  Further studies will investigate more detailed layouts and potential trade-offs between effective area and resolving power to maximize the figures of merit for the Arcus science case.

\subsection{CAT gratings and X-ray Surveyor}

The X-ray Surveyor is expected to have a mirror PSF of 0.5 arcsec half-power diameter (HPD).  Such a small PSF can translate into higher XGS resolving power as long as the gratings and the optical design do not degrade the angular resolution.  Initial ray-trace studies assuming 30 degree azimuthal sub-aperturing show that obtaining $R \sim$ 20,000 only requires very reasonable alignment tolerances similar to the case of our Arcus studies.  More details can be found in Ref.~10.

An X-ray Surveyor might employ one or more large-area retractable grating arrays as part of an XGS.  As in the case of the Chandra HETGS, the mass of such an array would be rather small, on the order of 10-20 kg, and easy to manipulate.

Since CAT gratings become highly transparent at energies above 2 keV one can continue to perform imaging with high energy resolution and minimal loss of effective area above 2 keV using a microcalorimeter, while at the same time performing high-resolution wavelength-dispersive soft x-ray spectroscopy, thereby maximizing science output during grating observations.

\subsection{Future Plans}

We are planning to move CAT grating technology forward in several directions.  Applying our fabrication approach to deeper gratings will lead to higher diffraction efficiencies.  Reducing the in-plane dimensions of L1 and L2 supports will increase effective area.  We plan to mount CAT gratings to thin machinable frames that can be mounted to a large grating array structure.  Standard machining tolerances are sufficient for design of and mounting to a GAS.  Only the roll degree-of-freedom requires an optical alignment technique, such as what was done for the Chandra HETGS.\cite{HETG}  We plan to populate a GAS with a small number of CAT gratings and perform efficiency and resolving power measurements before and after environmental testing (``shake-and-bake") to qualify for TRL5.

We will also continue to refine our ray-trace studies of relevant mission concepts.

\section{SUMMARY}

An x-ray grating spectrometer with large collecting area is the instrument of choice to move forward the science of high-resolution soft x-ray spectroscopy in astrophysics.  An XGS based on CAT gratings is a prime candidate for such an instrument, featuring high diffraction efficiency in high diffraction orders, relaxed alignment tolerances, low mass and resource demands, and synergy with imaging calorimeters.  These properties maximize effective area and resolving power and minimize instrument complexity, risk, and cost.

During the last year and a half we have experimentally demonstrated record-high resolving power and CAT grating efficiency, and extended CAT grating applicability to larger energies and blaze angles using atomic layer deposition.  CAT grating technology is on a clear path towards TRL5 and will soon be ready for inclusion in an XGS flight instrument.

\acknowledgments     
 
We gratefully acknowledge support from E.~Gullikson (ALS) and B. Chalifoux (SNL), and facilities support from the Nanostructures Laboratory and the Microsystems Technology Laboratories (both at MIT).  Special thanks go to W. Zhang, R. McClelland, K.-W. Chan, J. Niemeyer, and M. Schofield (all GSFC) for supplying the TDM.  A part of this work was performed at the Advanced Light Source at Lawrence Berkeley National Lab, which is supported by the Director, Office of Science, Office of Basic Energy Sciences, of the U.S. Department of Energy under Contract No. DE-AC02-05CH11231.  This work was supported by NASA grant NNX15AC43G.



\end{document}